\documentclass[12pt]{article}
\newsavebox{\ns}
\newsavebox{\dbrane}
\newsavebox{\dbshort}

\usepackage{epsfig}
\usepackage{amssymb}

\def\appendix{{\newpage\section*{Appendix}}\let\appendix\section%
        {\setcounter{section}{0}
        \gdef\thesection{\Alph{section}}}\section}

\newcommand\ba{\begin{eqnarray}}
\newcommand\ea{\end{eqnarray}}

\newcommand{\nn}{\nonumber}

\def\Dslash{\,\,{\raise.15ex\hbox{/}\mkern-12mu D}}
\def\Dbarslash{\,\,{\raise.15ex\hbox{/}\mkern-12mu {\bar D}}}
\def\delslash{\,\,{\raise.15ex\hbox{/}\mkern-9mu \partial}}
\def\delbarslash{\,\,{\raise.15ex\hbox{/}\mkern-9mu {\bar\partial}}}
\def\pslash{\,\,{\raise.15ex\hbox{/}\mkern-9mu p}}
\def\calDslash{\,\,{\raise.15ex\hbox{/}\mkern-12mu {\cal D}}}

\newcommand{\hh}{{1\over 2}}

\renewcommand{\ll}{_}
\newcommand{\uu}{^}
\newcommand{\pp}{\partial}
\renewcommand{\L}{\Lambda}
\renewcommand{\exp}[1]{{\rm exp}\{#1\}}
\renewcommand{\d}{\delta}
\newcommand{\m}{\mu}

\renewcommand{\m}{\mu}
\newcommand{\n}{\nu}
\newcommand{\s}{\sigma}

\newcommand{\G}{\Gamma}

\renewcommand{\r}{\rho}

\newcommand{\e}{\epsilon}

\newcommand{\sqd}{^2}

\renewcommand{\hh}{{1\over 2}}

\newcommand{\eee}[1]{\ba{#1}\ea}
\renewcommand{\th}{\theta}

\renewcommand{\b}{\beta}

\newcommand{\llsk}{\hskip .5in}

\newcommand{\st}{{}^*}
\newcommand{\D}{\Delta}

\newcommand{\lsq}{\left [}
\newcommand{\rsq}{\right ]}

\newcommand{\pr}{^\prime {}}

\newcommand{\apr}{{\alpha^\prime} {}}

\newcommand{\IZ}{\relax\ifmmode\mathchoice
{\hbox{\cmss Z\kern-.4em Z}}{\hbox{\cmss Z\kern-.4em Z}}
{\lower.9pt\hbox{\cmsss Z\kern-.4em Z}} {\lower1.2pt\hbox{\cmsss
Z\kern-.4em Z}}\else{\cmss Z\kern-.4em Z}\fi} \font\cmss=cmss10
\font\cmsss=cmss10 at 7pt
\newcommand{\inbar}{\,\vrule height1.5ex width.4pt depth0pt}
\newcommand{\IC}{{\relax\hbox{$\inbar\kern-.3em{\rm C}$}}}
\newcommand{\IQ}{{\relax\hbox{$\inbar\kern-.3em{\rm Q}$}}}
\newcommand{\IP}{\relax{\rm I\kern-.18em P}}

\renewcommand{\k}[1]{{k_{#1}}}

\newcommand{\Psb}{\bar{\Psi}}
\renewcommand{\k}{\kappa}
\renewcommand{\l}{{\tilde{\lambda}}}

\newcommand{\lrd}{\left (}
\newcommand{\rrd}{\right )}

\newcommand{\xx}{\tilde{\chi}}

\newcommand{\cm}{{\cal M}}

\newcommand{\ct}{\tilde{c}}

\renewcommand{\L}{\Lambda}
\renewcommand{\gg}{\nabla}
\renewcommand{\pr}{^\prime{}}

\newcommand{\IR}{\relax{\rm I\kern-.18em R}}
\def\blfootnote{\xdef\@thefnmark{}\@footnotetext}
\newcommand{\eeee}{&=&}

\newcommand{\hod}[1]{{\rm HO}^{+({#1})/}}

\newcommand{\bm}{\begin{matrix}}
\renewcommand{\em}{\end{matrix}}

\newcommand{\dn}{\Delta n}

\newcommand{\ee}[1]{\ba {#1} \ea}
\renewcommand{\xx}{{X^0{}}}

\newcommand{\hhc}{{1\over 2}\cdot}
\newcommand{\lno}{\left .}

\newcommand{\rba}{\right |}
\renewcommand{\dh}{\hat{\Delta}}

\newcommand{\up}[1]{^{({#1})}}
\newcommand{\lcb}{\left \{}
\newcommand{\rcb}{\right \}}

\newcommand{\tr}{{\rm tr}}
\newcommand{\phh}{{\hat{p}}}
\newcommand{\vws}{{V_{WS}}}

\newcommand{\vp}{\vec{\partial}}

\newcommand{\nnn}{\nonumber}

\newcommand{\otp}{{1\over{2\pi}}}

\newcommand{\qh}{\hat{{\cal Q}}}
\newcommand{\pso}{\psi^0{}}
\newcommand{\psx}{\psi^X{}}
\newcommand{\co}{{\cal O}}

\newcommand{\bbb}{\ba\begin{array}{c}}
\renewcommand{\eee}{\end{array}\ea}
\renewcommand{\eeee}{\nonumber\end{array}\ea}

\renewcommand{\cm}{{\cal D}}

\newcommand{\rws}{R_{WS}}
\newcommand{\lo}[1]{_{({#1})}}
\newcommand{\qe}{{Q^E}}
\newcommand{\psob}{\bar{\pso}}

\begin{document}

\begin{titlepage}

\begin{center}
~

~~~~~~~~~~~~~~~~~~~~~~~~~~~~~~~~~~~~~~~~~~~~~~~~~~~~~~~~~~~~~~~~~~~~~~~~~~~~~~~~~~~~~~~SU-ITP~04-35

\vskip 1.5 cm
{\large \bf Dynamical Dimension Change\\
\vskip 0.16cm
{\large \bf In}\\
\vskip 0.16cm
{\large \bf Supercritical String Theory}}

\vskip 1 cm
{Simeon Hellerman$^1$ and Xiao Liu$^{2}$}\\
\vskip 1cm

$^1${\sl School of Natural Sciences,
Institue for Advanced Study, \\
Princeton, NJ 08540 \\ {\tt simeon@ias.edu}\\}

\vskip .5cm

$^2${\sl Department of Physics and SLAC,
Stanford University, \\
Stanford, CA  94309 \\ {\tt liuxiao@itp.stanford.edu}}

\end{center}

\vskip 0.5 cm
\begin{abstract}
We study tree-level solutions to heterotic
string theory in which the number
of spacetime dimensions $10+n$ changes dynamically due to closed
string tachyon condensation. Taking the large-$n$ limit, we
compute the amount by which the dilaton gradient $\Phi_{,X^0}$
changes during the condensation process. At leading order in $n$
we find that the change in the invariant
magnitude of $\Phi_{,X^0}$
compensates the change in spacelike central charge on the
worldsheet, so that the total
worldsheet central charges are equal at
early and late times.  This result supports our interpretation of
the worldsheet theory as a CFT of
critical central charge which describes a dynamical reduction in
the total number of spacetime dimensions.
\end{abstract}

\end{titlepage}

\pagestyle{plain}
\setcounter{page}{1}
\newcounter{bean}
\baselineskip16pt
\tableofcontents


\newpage

\section{Introduction}
\label{intro}

The simplest tree-level solutions to superstring theory in
$D > 10$ are the \it timelike linear dilaton \rm
backgrounds.  For information on these, see \cite{Strominger},
\cite{Hellerman}, \cite{deAlwis}, \cite{Chamseddine},
\cite{Myers}.

These backgrounds posess
the maximal Poincar\'e symmetry of the supercritical
string, namely $D-1$ dimensional Poincar\'e symmetry of the
spacelike directions.  There is also an underlying $D$-dimensional
Poincar\'e symmetry spontaneously broken only by the dilaton
gradient, as pointed out by \cite{Myers} and elaborated in
\cite{deAlwis}, \cite{Chamseddine}.  Since the
time translation invariance is broken only by the dilaton,
there is a
time translation symmetry which acts on free string states
but is broken by interactions.

In \cite{Hellerman} a certain unstable heterotic string theory
in $10+n$ dimensions was introduced, and it was argued that
the ten-dimensional supersymmetric vacuum of SO(32) heterotic
string theory could be reached as an endpoint of tachyon
condensation in this theory.  The argument was based on
an approach to tachyon condensation which treated the
worldsheet theory classically.

Under a classical scale transformation, the embedding coordinate
$\xx$ shifts $\xx \to \xx + {\k Q}$, where $Q$ is the coefficient
in the worldsheet coupling \bbb \Delta\ll{(linear~dilaton)} {\cal
S} = {Q\over{4\pi}} \int~d\sqd\s ~\sqrt{g}~{\cal R}\up 2~\xx, \eee
where ${\cal R}\up 2$ is the worldsheet Ricci scalar. This
transformation leaves the action invariant up to terms independent
of the Liouville field $X^0$, and hence does not affect the
classical equations of motion.

Scale invariance is realized nonlinearly, in the sense that the
scale transformation of the set of classical fields has a
component which is zeroth order in the fields, as opposed to being
entirely first order in the fields, as in a usual CFT.  This
realization of conformal invariance has the consequence that one
can take an arbitrary scalar operator ${\cal O}\ll {(\Delta,
\Delta)}$ of arbitrary weight $(\D,\D)$ and make it into a
marginal operator by dressing it with an exponential
$\exp{2\b\xx}$, with the exponent $\b$ adjusted to give
exponential weight $(1 - \Delta, 1 - \Delta)$.  At the classical
level the exponent is determined by the equation\footnote{Note the
minus sign on the left hand side of this equation. This is
different from what is commonly seen in the literature. The
difference arises from the fact that the Liouville field in our
case is timelike as opposed to spacelike.} \bbb - Q\b = 1 -
\Delta.\eee

Quantum mechanically, the relationship is modified to \footnote{We
take $\alpha' =1$ in this paper. } \bbb - Q\b + \b\sqd = 1 -
\Delta. \eee Apart from this modification, the inclusion of
quantum effects has other consequences for the theory.

Most important from our point of view is that the
interpretation of the worldsheet theory as interpolating
between two linear dilaton vacua with different
dilaton gradients
$Q\ll\pm$ as $\xx\to \pm\infty$ depends on quantum effects for
consistency.  Specifically, in order for the total central
charge of the theory to be the same at $\xx \to + \infty$
as at $\xx \to - \infty$, the dilaton slope $Q$ must
change in order to compensate the loss of 'matter' central
charge due to the perturbation by the dressed relevant operator
$\co$.  This cannot happen at the classical level.  The aim
of this paper is to illustrate the mechanism by which
quantum effects renormalize the dilaton gradient $Q$ and
keep the total central charge of the theory constant
in both limits $\xx\to \pm\infty$.

\section{ A string worldsheet theory which
interpolates between two linear dilaton CFT's}

Let us make this more concrete by reviewing the specific theory
we will be analyzing.  In this section we will focus
on the classical analysis of the CFT.  From this point of
view it will be possible to see that the endpoint of closed
string tachyon condensation will be a reduced number of
spacetime dimensions.  We will also explain why it is
necessary to go beyond the classical analysis in order to see
the dynamical readjustment of the dilaton gradient in
target space time, which is
necessary for the consistency of the CFT.

\subsection{The type $\hod n$ string in a linear dilaton background}

We begin with a certain unstable heterotic string theory
introduced in \cite{Hellerman} and referred to as type $\hod n$.
The worldsheet theory is defined (in conformal gauge) by 9+n
spacelike embedding coordinates $X\uu i$ and their right-moving
worldsheet superpartners $\psi\uu i$, as well as a timelike
embedding coordinate $\xx$ and its right-moving superpartner
$\pso$.  There are also 32+n left-moving fermions $\l\uu a$.  All
the worldsheet fields are free.  To make the central charge
critical, we consider the theory in a timelike linear dilaton
background.  On a flat worldsheet, the only effect of the linear
dilaton is to add a term to the super-Virasoro generators which is
linear in $\xx,\pso$ rather than quadratic, and proportional to Q.
The condition for critical central charge $(\tilde{c},
c)\uu{matter} = (26,15)$ is \bbb 4 Q\sqd = n \eee

To preserve modular invariance we project onto states of
even overall worldsheet fermion number $F\ll W$.
With this projection there is a bulk tachyon $T\uu a$ living in the
fundamental representation of $SO(32+n)$\footnote{We use the term
'tachyon' here in the sense of \cite{Chamseddine}, to refer to a
field whose one-loop beta function equation is controlled by a
spacetime Lagrangian with a tachyonic mass term.  This is to be
distinguished from a 'tachyon' in the sense commonly used in the
literature on noncritical strings, which refers to a \it mode \rm
of a field which has negative \it frequency \rm squared in a
linear dilaton background.  We adopt the former terminology
because it is a more background-independent and Lorentz-invariant
usage than the latter, which should be more properly called an
'instability' than a 'tachyon'.}.

In order to stabilize the theory and relate it to conventional
heterotic string theory HO with gauge group $SO(32)$ in ten
dimensions, consider the $\hod n$ string on an orbifold
singularity, where the orbifold action is \bbb X\uu{ i} \to -
X\uu{i},\llsk\llsk\llsk i = 10,\cdots,9+n \cr \psi\uu{ i} \to -
\psi\uu{i},\llsk\llsk\llsk i = 10,\cdots,9+n \cr \l\uu a \to -
\l\uu a,\llsk\llsk\llsk a = 1,\cdots,32+n \eee The effect of this
orbifolding is to give the bulk tachyons Dirichlet boundary
conditions at the singularity \bbb T\uu a = 0~{\rm at}~X\uu {10} =
X\uu{11} = \cdots = X\uu{9+n} = 0 \eee

 The $+(n)$
referrs to the existence of $n$ extra spacetime dimensions, and
the diagonal slash $/$ refers to the fact that the fermion states
of the system are defined by projection onto states of even \it
overall \rm worldsheet fermion number $(-1)^{F_W} = +1$, rather
than even right- and left-moving worldsheet fermion numbers
separately, so the partition function for the fermions
is the \it diagonal \rm modular invariant.
For further details of the $\hod n$ string
theory and its relation to the critical $SO(32)$
heterotic superstring, see
\cite{Hellerman}.

\subsection{Tachyon perturbation}

\subsubsection{Classical analysis}

A tachyon background $T\uu a(X\uu\m)$ couples to the worldsheet
as a superspace integral
\bbb
\int ~d\th~\l\uu a~T\uu a(X\uu \m)
\eee
which in components amounts to a Yukawa coupling
\bbb
- i \l\uu a~T\uu a\ll{~~,\n}(X\uu\m)~\psi\uu\n
\eee
and a potential
\bbb
V(X\uu\m) = T\uu a (X\uu\m) T\uu a (X\uu\m)
\eee

In order to be scale invariant at the classical level, the tachyon
perturbation must satisfy \bbb -Q~T\ll{,\xx}\uu a (X\uu\m)  = T\uu
a(X\uu\m). \eee Since under classical scale transformations, the
$X\uu i$ are inert, and $X\uu 0$ shifts by $Q$, the equation above
is just a condition for a perturbation to be marginal. The
classical marginality equation is solved by \bbb T\uu a (X\uu\m) =
\exp{\b\xx} ~f\uu a (X\uu i), \eee with $\b =
- {1\over Q}$ and
$f\uu a$ an arbitrary set of functions of the spatial coordinates.

At the quantum level, the marginality
condition is modified to
\bbb
\eta\uu{\m\n}
T\ll{,X\uu\m X\uu\n}\uu a(X)
  - Q~T\ll{,\xx}\uu a (X)
= T\uu a(X)
\eee

For a tachyon profile of the form \bbb T\uu { 23+i}(X) = \m\uu i
R\ll i :\exp{\b\ll i \xx}~ \sin\lrd X\uu {i} / R\ll
i\rrd:,~~~~~~~~i=10,\cdots,9+n \cr\cr T\uu 1 = T\uu 2 = \cdots =
T\uu{32} = 0, \eee the condition is
\bbb - \b\ll i Q + \hh \b\ll i
\sqd = 1- {1\over{2R\ll i\sqd}}
\eee for each $i\in
\{10,\cdots,9+n\}$. In this paper we will simplify our discussion
by focusing on the limit where all the $R\ll i$ are large.  As
$R\ll i \to \infty$, we have \bbb \hh \b\ll i - {1\over{\b\ll i}} =
Q,~~~~~{\rm all}~i. \eee In this limit the worldsheet action for
the $X\uu i,\psi\uu i,$ and $\l\uu a$ is purely Gaussian, dressed
with $\xx$-dependence. The classical limit is the one in which we
take $Q\to\infty$, with $\b\ll i \to - {1\over Q}$.
We will fix our convention for the direction of time
by taking $Q < 0, \b > 0$ in which
the dilaton is decreasing with time and the tachyon is increasing.

In the classical limit, the physics of the worldsheet at $\xx\to
+\infty$ is clear: the worldsheet potential is \bbb V(X) = \sum\ll
i (\m\uu {9+i})\sqd~\exp{2\b \xx}~(X\uu{9+i})\sqd , \eee $\b =
- {1\over Q}$.  If we hold $\xx$ fixed and treat it as a background,
the effect of the potential is to give masses to the supercritical
embedding coordinates $X\uu i, i = 10,\cdots,9+n$, except for
those coordinates whose corresponding mass parameters $\m\uu i$
are exactly zero.  The nonzero masses become
infinitely large in the limit $\xx\to
+\infty$.  Likewise the fermions $\l\uu {23+i}$ and $\psi\uu{i}$
have large masses by virtue of their Yukawa couplings \bbb \m\uu i
\exp{\b\xx} \l\uu{23+i} \psi\uu i. \eee

So in the limit $\xx \to + \infty$, the massive fields
decouple in the Wilsonian sense, leaving the string
propagating in a smaller number of spacetime dimensions, and with
a smaller gauge group.  In particular if all of the $\m\uu i$
are nonzero, all $n$ of the extra embedding coordinates and their
superpartners, and all $n$ of the current algebra fermions $\l\uu a$
decouple.

\subsubsection{Classical limit versus central charge matching}

By treating the worldsheet
theory as classical this way,
it was shown in \cite{Hellerman} that the
worldsheet field content and discrete gauge charge
assignments in the limit $\xx\to+\infty$ are
those of the spacetime-supersymmetric heterotic string with
$SO(32)$ gauge symmetry.
One limitation of this treatment is that one cannot see
how the timelike dilaton gradient is reduced to zero
in the critical vacuum as a result of tachyon condensation.

Indeed, it is trivial to show that the effective worldsheet theory
at $\xx\to + \infty$ is \it exactly \rm the same as that of the
critical, spacetime-supersymmetric type HO string, modulo the
question of the linear dilaton term. But that is the best one can
do; at the level of classical field theory, eliminating the
massive degrees of freedom just means setting them to zero
classically.  There is no way that this can possibly affect the
linear dilaton coupling \bbb {Q\over{4\pi}} \int~d\sqd\s
~\sqrt{g}~{\cal R}\up 2~\xx. \eee

In order for the central charge to be constant, we would need
\bbb
Q\ll - \sqd - n = Q\ll +\sqd - (n - \Delta n),
\eee
where $Q\ll\pm$ is the limit of the dilaton gradient as
$\xx\to \pm\infty$, and
$\Delta n$ is the change in the number of spacetime dimensions,
which from the worldsheet point of view is equal to the number
of nonzero mass parameters $\m\uu i$.
At the classical level this central charge
matching condition automatically fails if $\Delta n \neq 0$,
since $Q\ll + = Q\ll -$ in the classical treatment.

\section{Semiclassical analysis of the worldsheet theory}

In this section we will discuss aspects of the
worldsheet theory considered at the semiclassical level.
We will explain the effect which gives rise to the
dynamical readjustment of the dilaton slope and show
some of the Feyman diagrams responsible for this effect.

\subsection{Dilaton slope renormalization by worldsheet
loop corrections}

We would like to go beyond the classical treatment of the
worldhseet theory, and see whether or not loop corrections to the
effective dilaton gradient can make the central charge matching
condition hold.  By 'loop corrections' and 'effective' we are
always referring to quantum effects on the worldsheet; we will
always be ignoring worldsheets of $g>0$ and working at tree level
in the spacetime sense.

\begin{figure}
        \begin{center}
                \includegraphics[-1.8in,1in][5in,2in]{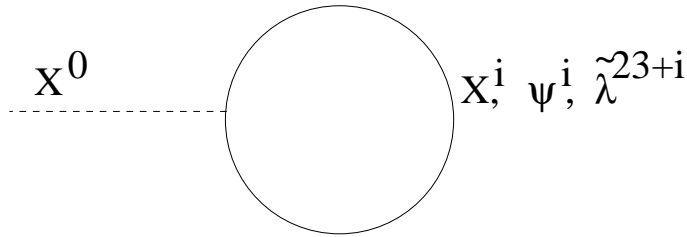}
        \end{center}
\vspace{1in}
         \caption{Integrating out massive degrees of freedom
gives rise to Feynman diagrams which renormalize the magnitude of
the coupling $\sqrt{g} ~{\cal R}\up 2~\xx$ in the worldsheet
Lagrangian.  In this diagram the solid lines are $X\uu i, \psi\uu
i, $ and $\l\uu{23+i}$ fields, and the dotted line is a $\xx$
field. }
        \end{figure}



Taking the limit in which the number $n$ of
extra spacetime dimensions becomes large,
we will find that the effect of integrating out
the massive embedding coordinates and fermions will be to renormalize
the effective dilaton gradient in such a way as to keep the
total central charge constant, despite the fact that
the central charge of the matter theory is reduced as
$X^0 \to \infty$.  This will be our main result.

It is easy to see that there are Feynman diagrams which would give
loop corrections to the effective dilaton slope.  Expanding the
potential term and Yukawa couplings to leading order in $\b$
yields trilinear interactions between two massive fields and an
$\xx$ field: \bbb {\cal L}\ll{o(\b\uu 1)} = \b \xx \sum\ll {i=
10}\uu{9+n} \lsq (\m\uu i)\sqd(X\uu i)\sqd - i \l\uu{23+i}\psi\uu
i \rsq \eee So one can construct Feynman diagrams where the
massive fields run in a loop, with an $\xx$ line attached.  In a
nonsupersymmetric theory this would give rise to a potential for
$\xx$, but the worldsheet supersymmetry guarantees that the
potential for $\xx$ must vanish, as we will calculate explicitly
later. The argument for this is simple: any Lorentz-invariant,
supersymmetric effective potential term for $\xx$ must be of the
form \bbb \int~d\th~\l\uu a~f\uu a (\xx) \eee But such a term
would violate the discrete worldsheet gauge symmetry under which
all the $\l\uu a$ are odd.  Therefore any effective potential
would have to have another odd field in it, such as one of the
$X\uu i$ with $i= 10,\cdots,9+n$, to compensate.

This argument holds only on a flat worldsheet.  On the sphere, the
curvature of the fiducial metric (which we will take to be the
round metric of radius $\rws$) breaks worldsheet supersymmetry.
Since the limit $\rws\to\infty$ is controlled by worldsheet
supersymmetry, the term of order $R_{WS}^2$ in the effective
action must vanish, but terms subleading in the volume of the
worldsheet are still allowed.  In particular it is possible to
generate a $\xx$-dependent effective term which scales as $\rws\uu
0$, which can be written as \bbb {1\over{4\pi}}
\int ~d\sqd \s~\sqrt{g}~{\cal
R}\up 2~\Phi\ll{1-loop}(\xx). \eee This term scales as $\rws\uu 0$
because the $\rws\sqd$ in $\sqrt{g}$ is cancelled by the
$\rws\uu{-2}$ in ${\cal R}\up 2$.

In other words, the spacetime dilaton gets a direct renormalization
from integrating out the massive fields, and that contribution
to the spacetime dilaton can be extracted by fixing
a large, positive background value of $\xx$, and extracting the
$\rws\uu 0$ term in the effective action
on the sphere.

\subsection{Wavefunction renormalization of the $\xx$ field}

We will find that there is a wavefunction renormalization of
the $\xx$ coordinate which will also be important.  Unlike
the dilaton coupling, this is an 'extensive' contribution
to the effective action, by which we mean that it scales
as $\rws\sqd$ rather than $\rws\uu 0$.  It cannot be
extracted from an effective action evaluated for static
$\xx$, so we must consider a background with
infinitesimal nonstatic perturbations added to a static piece,
in order to calculate it.

\begin{figure}
        \begin{center}
        \includegraphics[-1.3in,1in][5in,2in]{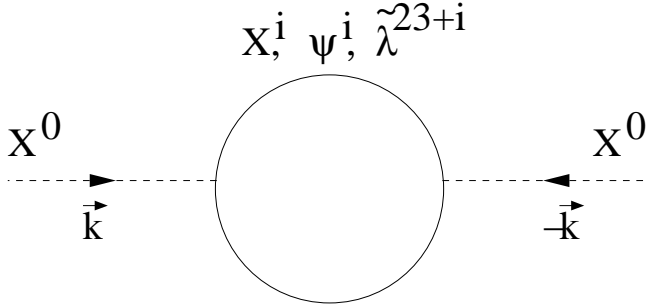}
        \end{center}
\vspace{1in}
         \caption{Integrating out massive fields
renormalizes the
kinetic term for the $\xx$ field.
}
        \end{figure}

We have not yet calculated the effective action $S\ll{eff}[\xx,
\pso]$ for the massless fields, but let us suppose we knew it,
and extract the effective kinetic term for $\xx$ from the
effective action.  The most general effective action for
$\xx$ allowed by the symmetries can be written in the form
\bbb
S\ll{eff}[\xx, \pso] = \int~d\sqd\s~
\k(\xx)(\vec{\partial}
\xx)\sqd +{\rm higher~derivatives} ~+\pso~{\rm terms}.
\eee
(The term with zero derivatives vanishes due to supersymmetry,
as explained earlier.)  To extract the kinetic
term, expand around the static background $\xx = \xx\lo 0$
to second order in a plane-wave fluctuation.  That is,
let $\xx(\s)$ be given by
\bbb
\xx = \xx\lo 0 + \e~\exp{i \vec{k}\cdot \vec{\s}}
+ \e\st~ \exp{-i \vec{k}\cdot \vec{\s}},
\eee
where $\e$ is an infinitesimal parameter.  Setting the
background values of the fermions $\pso(\s)$ to zero and
expanding
in both $k$ and in $\e,\e\st$, the effective action is
given by
\bbb
S\ll{eff}
[\xx\lo 0 + \e~\exp{i \vec{k}\cdot \vec{\s}}
+ \e\st~ \exp{-i \vec{k}\cdot \vec{\s}}, 0]
= 2 \k(\xx)~\vws~|\e|\sqd \vec{k}\sqd + o(|\e|\uu 3)
+ o(k\uu 4)
\eee
where $\vws$ is the volume of the worldsheet.  So if we are
given an effective action, we can extract $\k(\xx)
= 4\pi G\ll{\xx\xx}$ as
\bbb
\lno
\k(\xx\lo 0) = {1\over{2\vws}}{\d\over{\d\e}}{\d\over{\d\e\st}}
{\d\over{\d k\sqd}}  S\ll{eff}
[\xx\lo 0 + \e~\exp{i \vec{k}\cdot \vec{\s}}
+ \e\st~ \exp{-i \vec{k}\cdot \vec{\s}}, 0]
~~ \rba\ll{\e = \vec{k} = 0}
\cr\cr
~~~~~~~~~~~~~~~
=
\lno
{1\over{8\vws}}{\d\over{\d\e}}{\d\over{\d\e\st}}
{\d\over{\d k\uu a }} {\d\over{\d k\uu a }}  S\ll{eff}
[\xx\lo 0 + \e~\exp{i \vec{k}\cdot \vec{\s}}
+ \e\st~ \exp{-i \vec{k}\cdot \vec{\s}}, 0]~~\rba
\ll{\e = \vec{k} = 0}
\eee
\section{Large-$n$ limit}

\subsection{Central charge matching condition in the large-n
limit}

We will consider the limit in which the total number
of spacetime dimensions $10+n$ is large, and the dynamical
reduction in the number of spacetime dimensions is $\dn
\sim o(1)$.  In fact we will begin by taking $\Delta n = 1$;
the extension to higher $\Delta n$ is trivial
as long as $\Delta n / n$ is small.

In order to know what we would like the answer to be, let us expand
the central charge matching condition to leading order
in $n$.  Since $Q\ll -$ is the value of the dilaton
gradient which appears in the bare Lagrangian,
we will use $Q$ and $Q\ll -$ interchangeably:
\bbb
Q \equiv Q\ll - = - \hh \sqrt{n}.
\eee
In order for the central charge matching condition to
hold in the large-$n$ limit, the effective dilaton
slope $Q\ll +$ at large $\xx$ should be
\bbb
Q\ll + = - \hh\sqrt{n - \Delta n} \sim Q + {1\over 4}
{{\Delta n}\over{\sqrt{n}}} + o\lrd {{(\Delta n)\sqd}\over
{n\uu{3\over 2}} } \rrd.
\eee

\subsection{Feynman diagrams in the large-$n$ limit}

Let us now determine which Feynman diagrams we will need to
retain in order to compute the leading nontrivial correction
to the dilaton gradient in the large-$n$ limit.

\begin{itemize}

\item The direct contribution to the dilaton gradient is
a Feynman diagram with one external $\xx$ line.  The one-loop
contribution to this diagram has a single internal loop of massive
matter connected to the external line by a trilinear interaction
vertex, whose
 scales as $\b \sim - {1\over Q} \sim 2n\uu{-\hh}
+ o(n\uu{-{3\over 2}})$ in the large-$n$ limit.  Contributions
with more than one loop are suppressed by higher powers of $\b$
and are subleading in the large-$n$ limit.

\item The wavefunction renormalization for $\xx$ comes from
Feynman diagrams with two external $\xx$ legs.  Again the leading
contribution is a one-loop diagram with a massive internal loop
connected to the external lines by trilinear vertices.  This
diagram is of order $\b\sqd \sim 4 n\uu{-1}$ in the large-$n$
limit.  Redefining the $\xx$ field to make the effective kinetic
term canonical will contribute an amount of order $\b\sqd Q =
o(\b) = o(n\uu{-\hh})$ to the effective dilaton gradient in
canonical target space coordinates. Since this effect is of the
same order as the leading direct renormalization of $Q$, it must
also be retained.  Higher-loop kinetic term renormalizations
contribute $o(\b\uu 3 Q) = o(\b\sqd) = o(n\uu{-1})$ or less to the
canonical dilaton gradient in the large-$n$ limit, so we ignore
them.

\item The one-loop diagrams with two external $\xx$
lines and a quartic vertex have the same $\b$-scaling
(and hence $n$-scaling) as the diagram with two trilinear
vertices, but such diagrams
do not contribute to the renormalization of the kinetic
term;
the kinematics of the diagram
make the loop integral independent of the external momentum
$\vec{k}$.

\item As mentioned earlier,
we will also restrict ourselves to the approximation in which the
spatial wavelength of the tachyon in the supercritical dimension
is infinitely long.  In this approximation, the leading-order
worldsheet potential perturbation is purely quadratic as a
function of the massive fields $X\uu{i}, \psi\uu i, $ and
$\l\uu{23 + i}$. As a result, every interaction vertex has exactly
two massive lines.

\item If they existed,
one-loop diagrams without massive lines would
make contributions of the same order in $\b$ as the
diagrams we consider.  But every tree-level interaction vertex
has exactly two massive lines, so every loop diagram must contain
at least one massive loop. For one-loop diagrams with
only massless external lines,
massless lines cannot appear internally.

\item The noncontribution of massless internal lines at one
loop means that we do not need to distinguish between 1PI
and Wilsonian effective actions at leading order in $n$.
When all internal lines are massive, the only
effect of
leaving internal modes
with momenta below $E\ll{IR}$ fixed and integrating over them
is a set of contributions to the effective action which scale
with positive powers of $E\ll{IR} / M\ll{internal}$.  Since
$M\ll{internal}$ is proportional to $\m~\exp{\b \xx\lo 0}$,
such corrections vanish in the limit $\xx\lo 0 \to +\infty$
in which we will evaluate the effective action.

\item Since the $\xx$ and $\pso$ degrees of freedom are
essentially nondynamical for the purposes of the one-loop
calculation we will do, the result will be independent of
the details of operator ordering of the exponential
$\exp{\b\xx}$ appearing in the superspace integrand.  Beyond
one loop, the operator ordering of the exponentials
inevitably becomes important.

\item The operator ordering of the $(X\uu i)\sqd$ terms does
matter at one loop, but there is no ambiguity: the ordering is
determined by supersymmetry.  In particular, the $(X\uu i)\sqd$
potential is a simple product, not normal-ordered, because it must
be the square of the superpotential, which is linear in $X\uu i$.
The singularity of the product of coincident fields is an
exponential $\exp{2\b \xx}$ with UV divergent coefficient. This
singular term is cancelled by a second-order contact term between
two insertions of the Yukawa coupling $\exp{\b\xx} \l\uu{23 + i}
\psi\uu i$.  This is just the standard cancellation of the
one-loop vacuum energy in a supersymmetric system.

\end{itemize}

\section{One-loop effects in the worldsheet theory}

\subsection{Comments on signatures}

Since the kinetic term for the $\xx$ field is negative,
the worldsheet action in Lorentzian signature clearly does not define
a unitary quantum field theory.  This is not bad in and of
itself; ultimately we want to consider our 2D quantum theory as
part of a string theory, in which the super-Virasoro constraits
will eliminate the $\xx$ field as an independent degree of freedom,
just as in critical string theory in a static background.

However since the nonunitary $\xx$ sector is in our case an
interacting rather than a free theory, it is no longer clear
that the path integral suffices to define it

Following \cite{Strominger} we would like to define our theories by
Wick rotation from Euclidean space.  In doing so it is necessary
to assign the coordinate $X\uu 0$ a minus sign under
\it worldsheet \rm time-reversal transformations.  With this
assignment, $X\uu 0$ picks up a factor of $i$ under
Wick rotation, and the Euclidean worldsheet action
has positive definite kinetic term.

The result of this assignment is that the exponent $b$ appearing
in the tachyon perturbation of the Euclidean
worldsheet theory will be purely imaginary.  The screening
charge $\qe$ appearing in the Euclidean action is
imaginary as well.  They satisfy the relation
\bbb
\hh b + {1\over b}\lrd 1 - {1\over{2R\sqd}} \rrd  = \qe ,
\eee
which goes to
\bbb
\hh b + {1\over b} = \qe
\eee
in the long-wavelength limit $R\to\infty$.
Note that this is
the usual relationship between the 'screening
charge' $\qe$ and the Liouville exponent $b$ for unitary
super-Liouville theory.  (Here, $\xx$ plays the role of the Liouville
field.)\footnote{For bosonic Liouville theory, the usual
relationship is $b + b\uu{-1} = \qe$.  The difference has to do
with the anomalous dimensions of normal-ordered exponentials --
the dimension of $\exp{2b\xx}$ differs from twice the dimension
of $\exp{b\xx}$ by an amount of order $b\sqd$.  This means
that the bosonic potential term is not exactly marginal --
a potentially alarming problem!  At the order in
$b$ to which we are working, the effect of the non-marginality
of the potential term is cancelled exactly by the effect of
the contact term generated by two Yukawa coupling insertions, and the
theory is conformal to at least one loop.
The same apparent problem, and
the same resolution, arise in the $(1,1)$ super-Liouville theory.}

In the case of unitary Liouville theory
(and the related sine-Liouville and super-sine-Liouville theories)
$b$ and $\qe$ are real, whereas
in the action obtained by Wick rotating the Lorentzian worldsheet
theory, $b$ and $\qe$ are imaginary.

For our purposes we need not distinguish between the
two cases, since we are working in perturbation theory.
Indeed, amplitudes at fixed orders of perturbation
theory are polynomial in $b$, we can treat arbitrary
complex values of $b$ and $\qe$ uniformly, always
in the limit $|b|\to 0$.
Doing so begs the question of whether or not
the perturbative amplitudes we compute actually represent an
asymptotic expansion to any exact
amplitude in a  nonperturbatively well-defined theory.  Though
important, this question must await an \it a priori \rm
nonperturbative definition
of the theory, which we will not attempt here.

\subsection{The action of (0,1) super-sine-Liouville theory}

Since the massive fields couple to one another only through
the $\xx,\pso$ multiplet, it is clear that the one-loop
amplitudes we consider are simply proportional to
$\Delta n$, the number of massive fields.  This need not
be equal to $n$.  In fact, we must keep $\Delta n = o(1)$
in order for all higher loops to be suppressed relative to
the one-loop amplitudes.

Therefore we lose nothing by setting $\Delta n = 1$.  We will
denote the massive boson, its superpartner, and the
corresponding real current algebra fermion by $X, \psx,$
and $\l$, respectively.  The unique nonzero mass parameter,
we shall denote by $\m$.

The worldsheet action we are considering is the minimal
$(0,1)$ supersymmetric extension of the bosonic theory
known as sine-Liouville theory.  While the minimal
$(1,1)$ and $(2,2)$ supersymmetric versions of Liouville
and sine-Liouville theory have been studied extensively,
the minimal sine-Liouville theory with $(0,1)$ SUSY has
gone unexplored.  We now write down the Lorentzian and
Euclidean actions for these theories.

\subsubsection{Lorentzian action}

We consider the minimal extension of the theory which
admits right-moving (in Lorentzian signature) supercharge
$\qh\ll + = \qh$.  Then we define three SUSY multiplets
as follows:
\ba\begin{array}{c}
\qh \xx = - {i\over{\sqrt{2}}}
 \pso,\llsk  \llsk \qh\pso = \sqrt{2}~\pp\ll + \xx
\\\\
\qh X = - {i\over{\sqrt{2}}}
 \psx,\llsk  \llsk  \qh\psx = \sqrt{2}~\pp\ll + X
\\\\
\qh\l = F \llsk  \llsk \qh F = - i \pp\ll + \l
\end{array}\ea

We can create a Lagrangian which is supersymmetric on
a flat worldsheet up
to a total derivative by acting on a 'superspace' integrand
${\cal I}$ with $\qh$ to get a lagrangian:
\ba\begin{array}{c}
{\cal I} \equiv - {{\sqrt{2}}\over{4\pi}}
\pso \pp\ll - \xx
+
{{\sqrt{2}}\over{4\pi}}  \psx \pp\ll - X + {1\over{4\pi}}\l F
+ {{\m}\over{2\pi\sqrt{2}}} R~
\l~\exp{\b \xx} \sin(X / R)
\nnn\\\\
{\cal L}\ll{+ -} \equiv \qh\cdot {\cal I}
\nnn\\\\
= - \otp (\pp\ll + \xx)(\pp\ll - \xx) + \otp (\pp\ll + X)
(\pp \ll - X)  + {1\over{4\pi}}F\sqd \nnn \\\\
- {i\over{4\pi}} \pso\pp\ll - \pso
 +
{i\over{4\pi}} \psx \pp\ll - \psx
+{i\over{4\pi}} \l\pp\ll + \l\nnn\\\\
+
 {{\m}\over{2\pi\sqrt{2}}}  R F \exp{\b \xx}
\sin (X / R)
\\\\
 + {{i\m R \b}\over{4\pi}} \l \pso \exp{\b \xx}
\sin(X / R)
 + {{i\m }\over{4\pi}} \l \psx \exp{\b \xx} \cos (X / R)
\nn
\end{array}\ea
where ${\cal L}\ll{+-}$ is the Lagrangian which gets
integrated over $d\s\uu + d\s\uu -$.  Integrating out
the auxiliary field $F$ gives the lagrangian
\ba\begin{array}{c}
{\cal L}\ll{+ -}
= - \otp (\pp\ll + \xx)(\pp\ll - \xx) + \otp (\pp\ll + X)
(\pp \ll - X)   \nnn \\\\
- {i\over{4\pi}} \pso\pp\ll - \pso
 +
{i\over{4\pi}} \psx \pp\ll - \psx
+{i\over{4\pi}} \l\pp\ll + \l\nnn\\\\
 -
 {{\m\sqd  R\sqd}\over{8\pi}}  \exp{2 \b \xx}
\sin\sqd (X / R)
\\\\
 + {{i\m R \b}\over{4\pi}} \l \pso \exp{\b \xx}
\sin(X / R)
 + {{i\m }\over{4\pi}} \l \psx \exp{\b \xx} \cos (X / R)
\nn
\end{array}\ea

Note that there is no factor of $\b\sqd$ in front of the bosonic
potential.

Now we rewrite the action in the usual coordinates.  That
is, $\s\uu\pm \equiv \s\uu 0 \pm \s\uu 1$, so
$\pp\ll\pm = \hh(\pp\ll 0 \pm \pp\ll 1)$.  Further,
${\cal L}\ll{\s\uu 0 \s\uu 1} = 2 {\cal L}\ll{\pm}$, so
\ba\begin{array}{c}
{\cal L}\ll{\s\uu 0 \s\uu 1} =
 {{\eta\uu{ab}}\over{4\pi}}
 \lsq - (\pp\ll a \xx)(\pp\ll b\xx) + (\pp\ll a X)(\pp\ll b X)
  \rsq \nnn \\\\
+ {i\over{4\pi}} \lsq - \pso(\pp\ll 0 - \pp\ll 1) \pso
 +  \psx (\pp\ll 0 + \pp\ll 1) \psx
+ \l(\pp\ll 0 + \pp\ll 1) \l \rsq \nnn \\\\
 - {{\m\sqd}\over{4\pi}} R\sqd  \exp{2 \b \xx}
\sin\sqd (X / R)
\\\\
 + {{ i \m R \b}\over{2\pi}} \l \pso \exp{\b \xx}
\sin(X / R)
 + {{ i\m }\over{2\pi}} \l \psx \exp{\b \xx} \cos (X / R)
\nn
\end{array}\ea

For purposes of evaluating the massive fermion propagator,
it is convenient to package the two Majorana-Weyl fermions
$\psx$ and $\l$ into a single nonchiral Majorana fermion
$\Psi \equiv \lrd \bm{ \psx \cr \l} \em\rrd$.  We also
abuse notation and use $\pso$ to denote the a two-component form
of the Majorana-Weyl spinor $\pso\uu{\rm (one~component)}$ in which
the right-moving component is constrained to vanish: from
now on,
\bbb
\pso\equiv \pso\uu{\rm (two~component)}
= \lrd \bm {  \pso\uu{\rm (one~component)} \cr 0} \em \rrd.
\eee
Then we can write the action as
\ba\begin{array}{c}
{\cal L}\ll{\s\uu 0 \s\uu 1} =
 {{\eta\uu{ab}}\over{4\pi}}
 \lsq - (\pp\ll a \xx)(\pp\ll b\xx) + (\pp\ll a X)(\pp\ll b X)
  \rsq \nnn \\\\
+ {i\over{4\pi}} \lsq - \psob\G\uu a \pp\ll a \pso
 +  \Psb \G\uu a\pp\ll a \Psi
 \rsq \nnn \\\\
 - {{\m\sqd}\over{4\pi}} R\sqd  \exp{2 \b \xx}
\sin\sqd (X / R)
\\\\
 + {{i\m R\b}\over{2\pi}} \Psb \pso \exp{\b \xx}
\sin(X / R)
 + {{ i\m }\over{2\pi}} \Psb \Psi
 \exp{\b \xx} \cos (X / R),
\nn
\end{array}\ea
where
\bbb
\Psb \equiv - \Psi\uu T \G\uu 0
\eee
and we use the basis
\bbb
\G\uu 0 \equiv i \s\uu 2 = \lrd\bm{ 0 & 1 \cr -1 & 0 } \em\rrd
,~~~~~~~~~~~~~~~\G\uu 1 \equiv \s\uu 1
= \lrd\bm{ 0 & 1 \cr 1 & 0 } \em\rrd
\eee

Evaluated in the long-wavelength limit $R\to \infty$, the action
becomes
\ba\begin{array}{c}
{\cal L}\ll{\s\uu 0 \s\uu 1} =
 {{\eta\uu{ab}}\over{4\pi}}
 \lsq - (\pp\ll a \xx)(\pp\ll b\xx) + (\pp\ll a X)(\pp\ll b X)
  \rsq \nnn \\\\
+ {i\over{4\pi}} \lsq - \psob\G\uu a \pp\ll a \pso
 +  \Psb \G\uu a\pp\ll a \Psi
 \rsq \nnn \\\\
 - {{\m\sqd}\over{4\pi}}   \exp{2 \b \xx} X\sqd
\\\\
 + {{ i \m  \b}\over{2\pi}} \Psb \pso \exp{\b \xx} X
 + {{ i\m }\over{2\pi}} \Psb \Psi
 \exp{\b \xx} .
\nn
\end{array}\ea

\subsubsection{Euclidean action}

In order to Wick rotate, we send $\pp\ll 0 \to i\pp\ll 0$
and $\Psb \to \Psi \uu T C\st$.  As discussed earlier we also
perform a change of dynamical variables $\xx\to i \xx$.
>From this we obtain
\bbb
{\cal L}\uu E =
 {{\d\uu{ab}}\over{4\pi}}
 \lsq  (\pp\ll a \xx)(\pp\ll b\xx) + (\pp\ll a X)(\pp\ll b X)
  \rsq \nnn \\\\
+ {i\over{4\pi}} \lsq i \pso\uu {\rm T} C\st\G\uu a \pp\ll a \pso
 +  \Psi\uu{\rm T} C\st \G\uu a\pp\ll a \Psi
 \rsq \nnn \\\\
 + {{\m\sqd}\over{4\pi}}   \exp{2 b \xx} X\sqd
\\\\
 + {{ i  \b\m}\over{2\pi}} \Psi \uu{\rm T} C\st \pso \exp{b \xx} X
 + {{ i\m }\over{2\pi}} \Psi\uu{\rm T} C\st \Psi
 \exp{b \xx} .
\eee

On a curved worldsheet, change the partial derivatives into
covariant derivatives, and add the dilaton coupling:
\bbb
{\cal L}\uu E =
 {{g\uu{ab}}\over{4\pi}}
 \lsq  (\pp\ll a \xx)(\pp\ll b\xx) + (\pp\ll a X)(\pp\ll b X)
  \rsq
+ {i\over{4\pi}} \lsq i \pso\uu {\rm T} C\st\G\uu a \gg\ll a \pso
 +  \Psi\uu{\rm T} C\st \G\uu a\gg\ll a \Psi
 \rsq \nnn \\\\
 + {{\m\sqd}\over{4\pi}}   \exp{2 b \xx} X\sqd
 + {{ i  \b\m}\over{2\pi}} \Psi \uu{\rm T} C\st \pso \exp{b \xx} X
 + {{ i\m }\over{2\pi}} \Psi\uu{\rm T} C\st \Psi
 \exp{b \xx}
\cr\cr
+{{\qe}\over {4\pi}} \xx~{\cal R}\up 2
\eee
\subsection{The one loop effective action for $\xx$}

Computing the effective action for $\xx$ at one loop reduces to
the computation of a determinant depending on the background
configuration of $\xx$.

If the massive bosonic and fermionic degrees of freedom are
represented as $B\uu i$ and $f\uu i$, respectively, then
let their action in a given $\xx$ background be denoted
by
\bbb
{\cal S}\ll{massive} \equiv \hh
S\ll{(IJ)}B\uu I B\uu J + \hh A\ll{[ij]} f\uu i f\uu j,
\eee
where $S\ll{(IJ)}$ is a symmetric matrix and
$A\ll{[ij]}$ is an antisymmetric matrix, both depending on
the background configuration of $\xx$, not necessarily static,
and on the fiducial metric on the sphere, which we are
taking to be round of radius $\rws$.

The effective action for $X\uu 0$ is the logarithm of a Gaussian
integral, and is equal to \bbb (\Delta {\cal S})\ll{1-loop} =
c\ll{meas.} + {1\over 2} \ln\lsq {{\det (S)}\over{\det (A)}} \rsq
\eee $c\ll{meas.}$ is a constant involved in the definition of the
measure, which is independent of the worldsheet metric or any of
the dynamical variables such as $\xx$.
\subsection{Regulating the determinants}

The bosonic and fermionic contribution to the logarithm are
separately divergent.  Let us regulate the determinant in a
definite way to eliminate this problem.  Let $B(x) \geq 0 $ be a
positive semidefinite infinitely differentiable bump function
which interpolates between $B(0) = 1$ and $B(\infty) = 0$.  Let
$\L$ be a mass scale which we shall later take to infinity.  Then
let $\m\pr$ be an arbitrary mass scale, and substitute
\ba\begin{array}{c} S \to (\m\pr)\sqd \lrd S / (\m\pr)\sqd  \rrd
 \uu{B\lrd {{-\nabla\sqd}\over{\L\uu{2}}}\rrd}
\equiv S\ll\L
\\\\
A \to \m\pr \lrd A/\m\pr \rrd
 \uu{B\lrd {{-\nabla\sqd}\over{\L\uu{2}}}\rrd}
\equiv A \ll \L
\end{array}\ea
Raising $S$ and $A$ to the power $B(-\nabla\sqd / \L\sqd)$ has the
effect of damping the contributions of the eigenvalues greater
than $\L\sqd$. On a particular eigenvector of $S$ or $A$ the
regulated operator $S\ll \L,  A\ll \L$ will have an eigenvalue
which is damped: $\l \to \l\uu {B(\l / \L\uu{\D})}$, which
approaches $1$ as $\l\to\infty$ for fixed $\L$, but approaches
$\l$ as $\L \to\infty$ for fixed $\l$.  So in that non-uniform
sense, the damping exponent has no effect in the limit
$\L\to\infty$, but it regulates the determinant for any finite
$\L$.  (We choose $B(x)$ to fall off sufficiently quickly that the
regulated sums of
logarithms of eigenvalues of the
the operators $A,S$ converge for finite
$\L$.)

\subsection{Results}

\subsubsection{For $\xx = \xx\lo 0 = {\rm const}.$}

For a static background, we compute the bosonic and fermionic
contributions to the one-loop effective action.  Up to terms which
vanish as $\L\to \infty$, the results are \bbb {\cal S} \ll{bos} =
{1\over 2} \ln \det(S\ll \L) = \vws C(\L) + 2
\D\uu{}\Phi\ll{bos}(\xx) \cr\cr {\cal S} \ll{ferm} = - {1\over 2}
\ln~\det(A\ll \L) = - \vws
 C(\L)
+ 2 \D\uu{}\Phi\ll{ferm}(\xx)
 \cr\cr
{\rm with} \cr\cr C(\L)= {{\L\sqd}\over{8\pi}} \lrd 2 I\ll 1
\ln\lrd {{\L}\over{\mu\pr}} \rrd + I\ll 2  \rrd \cr\cr {\rm and}
\cr\cr \D\uu{}\Phi\ll{bos}(\xx) =  {1\over 6} b\xx +{1\over 6}
\ln\lrd {\m\over{\m\pr}} \rrd \cr\cr \D\uu{}\Phi\ll{ferm}(\xx) =
{1\over{12}}b\xx + \hh\ln\lrd {\m\over\L} \rrd - {5\over{12}}
\ln\lrd{\m\over{\m\pr}} \rrd - {1\over 4}\lrd I\ll 3 + I\ll 4
\rrd. \eee Here $I\ll{1,2,3,4}$ are finite, nonuniversal numerical
constants which depend on the precise form of the bump function,
which we list in the appendix for completeness.  The divergent
constant in $\D\uu{}\Phi\ll{ferm}$ is also nonuniversal and its
effect on the worldsheet partition function can be removed with a
local counterterm: \bbb \lrd {\D\cal L}\rrd\ll{counterterm}
 = - {{{\cal R}\up 2}\over{8\pi}} \ln\lrd
{\m\over\L} \rrd.
\eee
Likewise, the finite $\xx$-independent
terms can be absorbed into an
overall finite redefinition of the string coupling.  The only
physical information is the renormalization of the
dilaton slope coming
from the one-loop graph.  The total direct renormalization
contributed by bosonic and fermionic massive fields is:
\bbb
\D \uu{}\Phi =
 {1\over 4}b\xx + ({\rm const.})
\cr
\D\Phi\ll{,\xx} = {1\over 4} b
\eee
\subsubsection{Static background with perturbations}

Expanding around \bbb \xx = \xx\lo 0 + \e~\exp{ik\s} + \e\st\exp{-
ik\s}, \eee to second order in $k$ and second order in $\e,\e\st$
we find the effective action coming from a bosonic and a fermionic
loop of massive modes.  This computation encodes the
renormalization of the kinetic term $G\ll{\xx\xx}$ for the
Liouville field $\xx$: \bbb
  \lno \D\ll{bos} {\cal S}   \rba\ll{o(k\sqd \e\st\e)}
 = \lno \hh \ln \det(S\ll\L)  \rba\ll{o(k\sqd \e\st\e)}
\equiv {\vws\over{4\pi}} \lrd \D G\ll{\xx\xx} \rrd\ll{bos}
~\e\st\e~\vec{k}\sqd
\cr\cr
 \lno \D\ll{ferm} {\cal S}  \rba\ll{o(k\sqd \e\st\e)}
 = - \lno \hh \ln \det(A\ll\L)  \rba\ll{o(k\sqd \e\st\e)}
\equiv {\vws\over{4\pi}} \lrd \D G\ll{\xx\xx}\rrd\ll{ferm}
~\e\st\e~\vec{k}\sqd
\cr\cr
 \lrd \D G\ll{\xx\xx} \rrd\ll{bos}
 \equiv +{{b\sqd}\over 3}
\cr\cr
 \lrd \D G\ll{\xx\xx}\rrd\ll{ferm}
 \equiv +{{b\sqd}\over 6}
\eee
for a total renormalization of
\bbb
\lrd \D G\ll{\xx\xx} \rrd  = +{{b\sqd}\over{2}}
\eee
\subsection{Central charge in the small-$b$ limit}

Now we compute the one-loop renormalization of the
linear dilaton's contribution to the central charge.

\subsubsection{In the limit $\b\xx\to -\infty$}

As $\b\xx\to -\infty$ (equivalently $\m\to 0$)
the worldsheet theory is weakly coupled and
the central charge can be computed in free field theory. The
number of current algebra fermions is $32+n$ and the number of
spacetime coordinates and their right-moving superpartners is
$10+n$.  We can choose the magnitude of the linear dilaton to be
arbitrary, since the $\xx\to -\infty$ value of dilaton gradient is
the number $Q$ which appears in the bare action.  We also choose
the bare kinetic term for $\xx$ to be the canonical one
${1\over{4\pi}} G\uu{(bare)} \ll{\xx\xx}(\vp \xx )\sqd$ with
$G\ll{\xx\xx}\uu{(bare)} = 1$. The central charge contributed by a
dilaton gradient is \bbb c\ll{(linear~dilaton)}
 =6 G\ll{(bare)}\uu{\xx\xx}
\Phi\uu{(bare)}\ll{,\xx} \Phi\uu{(bare)}\ll{,\xx} = \ct
\ll{(linear~dilaton)}
\cr\cr = 6(\qe)\sqd \eee

\subsubsection{In the limit $\b\xx\to +\infty$}
In this limit the theory of $\xx,X,\pso,\l,\psx$
as a whole is strongly interacting, and
we cannot rely on the same free field calculation to determine the
central charge as we used in the limit $\b\xx\to - \infty$.
Nonetheless in the limit $\b\xx\to +\infty$ (equivalently $\m\to
\infty$)
the effective theory of $\xx,\pso$ obtained
by integrating out the massive fields $X,\psx,\l$, is a weakly
interacting one.

In the effective theory of $\xx,\pso$ at $\xx \to +\infty$, we
have \bbb \Phi\ll{,\xx} = \qe  + {1\over 4} b + o(b\sqd) \cr\cr
G\ll{\xx\xx} = 1 + {1\over 4} b\sqd + o(b\uu 3) \cr\cr G\uu{\xx\xx} =
1 - {1\over 4} b\sqd + o(b\uu 3) \eee

Working
as always in units where $\apr \equiv 1$, the standard formula
for the superstring is
\bbb
c\ll{(linear~dilaton)}
= 6 G\uu{\m\n} \Phi\ll{,\m} \Phi\ll{,\n} =
6\lsq (\qe)\sqd -{1\over 4}
 b\sqd  (\qe)\sqd
+ {1\over 2} b\qe\rsq + o(b\uu {1})
\eee
for a dilaton $\Phi$ which
is linear in the same set of coordinates $X\uu\m$
in which the string-frame
metric $G\ll{\m\n}$ is constant.  Since $Q\uu E$ is
of order $b\uu{-1}$ we must retain the $(Q\uu E)\sqd b\sqd$
and $(Q\uu E) b$ terms
for consistency.  The first of the two terms comes from the
correction to the metric and the second comes from the
direct renormalization of the dilaton gradient.

At leading order in $b$, the marginality condition for the tachyon
is
\bbb
\qe = {1\over b} + o(b),
\eee
so the change in the central charge of the linear dilaton
theory is
\bbb
\D c\ll{(linear~dilaton)} = + {6\over 4} b\qe + o(b)
= + {3\over 2} + o(b)
\eee

Since the change in the matter central charge is
$\D c\ll{matter} = - {3\over 2}$, the central charge remains
constant at the order in $b$ to which we have calculated.

\section{Spacetime interpretation}

In order to interpret our results, let us return to Lorentzian
signature.  To do this, let \bbb \qe\equiv i Q \cr b\equiv i \b
\cr \hh \b - {1\over{\b}} = Q \eee Upon simultaneous Wick rotation
of worldsheet and target space coordinates, we recover the usual
Lorentz-signature action, with a negative kinetic term for $\xx$,
timelike dilaton gradient $Q$ (which is negative) and tachyon
decay rate $\b$.

The worldsheet theory has two weakly coupled limits:
the first is $\xx\to -\infty$, in which the bare
potential vanishes.  The second is $\xx\to +\infty$,
in which the interaction term grows large, some degrees
of freedom decouple, and the resulting \it effective \rm
theory of $\xx, \pso$
is weakly coupled.  We can calculate the central charge
reliably in both regimes.  If our worldsheet theory really
is a CFT, the central charge must be a c-number, and the
two calculations must match.  Since we are only working
to one loop, it is only the $o(n\uu 0)$ change in the central
charge which we can compute accurately.  We find that the
total central charge of the theory
at $\xx\to +\infty$ is the same as that at $\xx \to - \infty$.

If the original theory has $n$ supercritical dimensions and real
current algebra fermions, the total central charge of the matter
theory is \bbb c = 15 + {{3n}\over 2} + 6 (\nabla\ll\m
\Phi)(\nabla\uu\m\Phi) \cr {\tilde{c}} = 26 + {{3n}\over 2} + 6
(\nabla\ll\m \Phi)(\nabla\uu\m\Phi) \eee  Since the gradient
of $\Phi$ lies in a timelike direction $\xx$, the linear
dilaton contribution to the central charge is negative.

The physical interpretation is that the theory we are studying
is a worldsheet CFT of a fundamental string,
which interpolates between two different linear dilaton
theories in two different directions in target space.

Quite apart from the issue of signatures, this interpretation
stands in contrast with the behavior of bosonic Liouville
and sine-Liouville theories, or any of their other
supersymmetric extensions.
In order to interpolate
between two linear dilaton theories, the
minimum of the worldsheet potential must lie at
zero in both infinite limits in Liouville field
space $\r\to\pm\infty$ (in our case, $\xx\to\pm\infty$).
But none of the other (sine-)Liouville theories has
that property.  In this sense, the $(0,1)$ super-sine-Liouville
theory is distinguished as a worldsheet theory with the
ability to describe dynamical transitions between
simple string vacua with different numbers of target space
dimensions.

\begin{figure}
        \begin{center}
        \includegraphics[2.5in,1in][3.5in,8.5in]{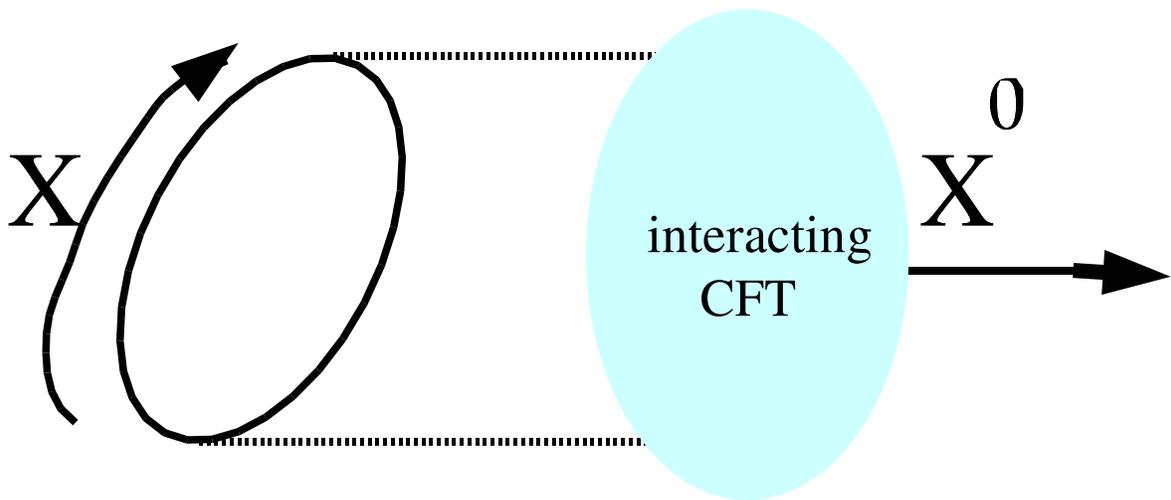}
        \end{center}
\vspace{-3in}
         \caption{Behavior of the worldsheet theory:
For large negative $\xx$, the worldsheet theory is
a free CFT of $10+n$ embedding coordinates whose dilaton
gradient has magnitude $Q =  - {\sqrt{n}\over 2}$.
at intermediate $\xx$ the worldsheet theory is
a nontrivial interacting CFT.  For large
positive $\xx$ it is a free CFT of $9+n$
embedding coordinates whose dilaton gradient
has magnitude $Q = - {\sqrt{n - 1}\over 2}$.  To order
$n\uu{-\hh}$, explicit calculation of the renormalization of
the dilaton gradient at $\xx\to\infty$ supports this
picture.  To make the figure as simple as possible, we have not
indicated the orbifolding of the coordinate $X$. }
        \end{figure}

The classical behavior of the 2D worldsheet theory motivated
this interpretation of the $(0,1)$ super-sine-Liouville action
\cite{Hellerman}.
The tachyon of the orbifolded
type $\hod n$ theory couples to the
heterotic string worldsheet as a mass term for one (or more)
of the coordinates transverse to the orbifold fixed locus.
As the tachyon grows with target space
time $\xx$, the mass term gets larger and pins the fundamental
string to the origin of the coordinate which is
being eliminated.  As a result, the number of spacetime
dimensions describing the string theory is reduced as
$\xx\to +\infty$.  In the limit $\xx = +\infty$ the worldsheet theory
is again free; it describes strings propagating in $10 + n - \Delta
n$ spacetime dimensions with gauge group $SO(32+n - \Delta n)$.

When superstrings propagate in flat space in a number of dimensions
other than ten, there must be a linear dilaton gradient
which compensates the central charge.  It is necessary for
the self-consistency of our interpretation that when the
number $n+10$ of spacetime dimensions changes, the magnitude of
the dilaton gradient must change to compensate, so that
the combination
\bbb
 {3\over 2} n - 6 Q\sqd
\eee
remains unchanged, and in particular stays equal to zero.

The dynamical readjustment of the dilaton gradient
is the sole feature of the expected spacetime background which
cannot be understood at the classical level on the
worldsheet.  Happily, we have found that the theory contains
quantum effects which supply the necessary
change in the dilaton gradient.  Physical expectations
for the change in the dilaton gradient agree
quantitatively with explicit calculation in the limit
$n\to \infty$.

It would be interesting to study the properties of our
theory beyond the one-loop level, including a nonperturbative
definition of the theory and a
proof of conformal invariance.
Liouville and sine-Liouville theories with
various amounts of supersymmetry have been constructed and solved
exactly via many different methods.  (See, for instance,
\cite{lt1}-\cite{lt9}.)  We anticipate that a further study of
the dynamics of $(0,1)$-supersymmetric
sine-Liouville theory will yield may interesting insights
into dynamical dimension change in string theory.

\bigskip
\centerline{\bf{Acknowledgements}} The authors are indebted to
many of their colleagues for helpful discussions,
particularly J. Maldacena, L. Susskind, G. Girib\'et, J.
M${^{\underline{\rm c}}}$Greevy, S. Kachru,
A. Pakman, S. Shenker, and E. Silverstein.
S.H. would like to thank the participants of the IAS
theory group meetings for comments and questions which did much to
motivate this research, and also the Stanford ITP and the
Perimeter Institute for hospitality while this work was in
progress. X.L. gratefully acknowledges
the Prospects in Theoretical Physics program
for hospitality during the course of this research.
The work of S.H. was supported by DOE Grant
DE-FG02-90ER40542.

\appendix{Direct renormalization of the dilaton gradient}

For purposes of computing 'extensive' terms, we can ignore
the finite volume and finite curvature, and just replace
the covariant derivatives with partial derivatives again.

This calculation will do little other than check that we have
chosen a correct, supersymmetric definition of the
measure.  Many things, including simple (independent)
rescalings of bosons and fermions in the same multiplet at
the classical level, can produce a nonsupersymmetric result
for the one-loop potential.  This can of course be cancelled
by a local counterterm, but it is most convenient to take a
definition of the measure which does not require this.

For the purpose of evaluating traces, we will make use of the
identity
\ee{
{\rm Tr} f(\hat{p}) = {{\vws}\over{(2\pi)\sqd}}
\int d\sqd p~f(p).
}

\subsection{Effective potential and effecitve linear dilaton}

The effective action for a scalar field $\xx$ which is
constant on a round worldsheet of radius $\rws$ has two
contributions: the first is a potential, which scales
with two powers $\rws\sqd$ of the worldsheet radius; the
second contribution is an effecitve coupling $f(\xx)
{\cal R}\uu{(2)}$ to the worldsheet Ricci scalar.  This
term is down by two powers of $\rws$ relative to the
potential term.

In order to extract the term subleading in $\rws$ in which
we are interested, we first focus on the leading, potential term.
Due to worldsheet supersymmetry, the contributions to the
effective potential term will vanish
due to cancellation between boson and fermion
loops.

To extract the effective potential from the effective action,
set the scalar $\xx$ to a fixed constant value and divide by
the volume $V\ll{WS} = 4\pi \rws\sqd$ of the worldsheet.
The effective action
is ${\hbar\over 2}$ times the logarithm
of the bosonic determinant minus the
logarithm of the
fermionic determinant. We will evaluate
each finite-volume loop diagram separately,
beginning with the bosonic loop.


\subsection{Determinants in finite volume}

\subsubsection{The bosonic contribution to the potential}

Consider a background $\xx(\s)
= \xx\lo 0$ independent of $\s$, and
let $M\equiv \m \exp{b\xx} = \m \exp{b\xx\lo 0}$.
The bosonic determinant is
\ee{
\sum\ll{n = 0}\uu\infty (2 n + 1) ~
\ln \lrd {{M\sqd}\over
{({\mu\pr})\sqd}} + {{n\sqd + n}
\over{\rws\sqd({\mu\pr})\sqd}}
 \rrd
}
We regulate the determinant, the result of which is that
we multiply each log by
\ee{
B\lrd {{n\sqd + n}\over{\rws\sqd \L\sqd}}\rrd
}
The regulated sum is
\ee{
\sum\ll{n = 0}\uu\infty (2 n + 1) ~\ln \lrd {{M\sqd}\over
{({\mu\pr})\sqd}} + {{n\sqd + n}
\over{\rws\sqd({\mu\pr})\sqd}}
 \rrd \cdot B\lrd {{n\sqd + n}\over{\rws\sqd \L\sqd}}\rrd
}
To evaluate this, let us use the Euler-Maclaurin formula
\ba\begin{array}{c}
\sum\ll{n = a}\uu b f(n) = \int\ll a \uu b dn ~ f(n) +
\hh(f(a) + f(b)) + {1\over{12}}\lrd f\pr(b) - f\pr(a) \rrd +
o(f\uu{\prime\prime\prime})
\\\\
= T\ll {-1} + T\ll 0 + T\ll 1 + {\rm remainder}
\end{array}\ea
\ba\begin{array}{c}
T\ll{-1} \equiv
\int\ll 0\uu\infty dn~ (2 n + 1) ~\ln \lrd {{M\sqd}\over
{({\mu\pr})\sqd}} + {{n\sqd + n}
\over{\rws\sqd({\mu\pr})\sqd}}
 \rrd \cdot B\lrd {{n\sqd + n}\over{\rws\sqd \L\sqd}}\rrd
\\\\
= \L\sqd \rws\sqd \int\ll 0\uu\infty du
\lsq ~\ln\lrd {{\L\sqd}\over{({\mu\pr})\sqd}}  \rrd +
\ln \lrd {{M\sqd}\over
{\L\sqd}} + u
 \rrd \rsq B(u)
\end{array}\ea

By taking $\L\sqd {d\over{d\L\sqd}}$ of the integral (without
the $R\sqd \L\sqd$ prefactor) we can see that the leading
divergence is a potential of the form
\ba\begin{array}{c}
V(M) = {\hbar\over 2} T\ll{-1}
= {\hbar\over{8\pi}} \lrd I\ll 1 \L\sqd \ln\lrd
{{\L\sqd} \over{({\mu\pr})\sqd}} \rrd + I\ll 2 \L\sqd \rrd
\end{array}\ea
where
\ee{
I\ll 1 \equiv \int \ll 0 \uu\infty du~B(u),\llsk\llsk\llsk
I\ll 2 \equiv \int \ll 0 \uu\infty du~B(u)~\ln u
}

Note that this 'potential' is actually a cosmological constant,
having no dependence on $M$.  It explicitly breaks scale invariance
(since it is a term coming from the regulator) and must
be subtracted off with a counterterm in order to restore scale
invariance if it is not cancelled by another effect, which
of course it is in the supersymmetric theory.

\subsubsection{The fermionic contribution to the effective
potential}

Now let's redo this for the fermionic determinant.  In some basis,
the operator ${\cal D}
\equiv \G\uu i \nabla\ll i $ is equal to~\cite{eigenvalues}
\ee{
{1\over \rws} \lsq \bm { i (n+1) & 0 \cr 0 & - i (n+1) } \em \rsq,
}
with multiplicity $2n+2$.

So the eigenvalues of ${\cal D} + M$ are equal to $M \pm {{i (n+1)}
\over \rws},n\geq 0$,
with multiplicity $2n+2$ each.

In order to take the regulator into account, we use the fact
that $(\G\uu i \nabla\ll i) \sqd = \nabla\uu i \nabla\ll i -
{1\over 4} {\cal R}\uu{(2)}$.  This is the 'Weitzenbock formula',
but it is straighforward to work out by hand.
For a spherical worldsheet
of radius $\rws$, the Ricci scalar is ${2\over{\rws\sqd}}$, so
\ee{
\nabla\sqd = {\cal D}\sqd + {1\over {2\rws\sqd}} =
{{\hh - (n+1)\sqd}\over{\rws\sqd}}
}

So the log of the regulated determinant is
\ba\begin{array}{c}
\sum\ll{n = 0}\uu\infty ~(2n + 2) ~\ln \lrd
{{M\sqd}\over{({\mu\pr}) \sqd}}
 + {{(n+1)\sqd} \over{\rws\sqd({\mu\pr})\sqd}}
 \rrd ~B\lrd {{(n+1)\sqd - \hh}\over{\rws\sqd\L\sqd}} \rrd
\end{array}\ea

\subsubsection{The extensive term}

First we extract the term $T\ll {-1}$ proportional to volume.

This term is
\ba\begin{array}{c}
T\ll{-1}\equiv \int\ll 0 \uu\infty dn~(2n + 2) ~\ln \lrd
{{M\sqd}\over{({\mu\pr}) \sqd}}
 + {{(n+1)\sqd} \over{\rws\sqd({\mu\pr})\sqd}}
 \rrd ~B\lrd {{(n+1)\sqd - \hh}\over{\rws\sqd\L\sqd}} \rrd  \nn\\\\
= \L\sqd \rws \sqd \int\ll 0 \uu\infty du ~\lcb
\ln \lrd {{\L\sqd}\over{({\mu\pr})\sqd}} \rrd +
\ln \lrd
{{M\sqd}\over{\L \sqd}}
 + { 1 \over{\rws\sqd\L \sqd}}
+ u  \rrd
 \rcb ~B\lrd
 u
+ {{1 }\over{2 \rws\sqd\L\sqd}}
 \rrd \nn\\\\
\end{array}\ea
where we have used the changes of variables $t \equiv
(n+1)\sqd - \hh$ and $u \equiv {t\over{\L\sqd \rws\sqd}}$.

To leading order in $\rws\sqd \L\sqd$, the integrand is
the same as in the log of the bosonic determinant.  The
difference is of subleading order in $\rws\sqd\L\sqd$ and
it is equal to
\ba\begin{array}{c}
T\ll{-1}\uu{fermionic} - T\ll{-1}\uu{bosonic}~~~\nn \\\\
\simeq \L\sqd \rws\sqd \int\ll 0 \uu\infty du~
{1\over{\L\sqd \rws\sqd}}
{1\over{{{M\sqd}\over{\L \sqd}}
+ u }} B(u) +  \ln \lrd
{{M\sqd}\over{\L \sqd}}
+ u  \rrd  {1\over{2\L\sqd \rws\sqd}} B\pr (u)~~~ ,
\end{array}\ea
with a remainder which vanishes as $\L\sqd \to\infty$ with
the other quantities held fixed.  The first term in the
integrand was obtained by expanding the logarithm in
${1\over{\L\sqd \rws\sqd}}$, the second term by expanding the
regulator function in ${1\over{\L\sqd \rws\sqd}}$.

First, we point out that
in the limit $\L\sqd \rws\sqd \to\infty$, the second
term asymptotes to a constant, and does not behave as
a logarithm of $M$.  This
is useful to us, as it means that the renormalization
of the dilaton gradient will not depend on the form of the
regulator function, in accordance with the field-theoretic
principle of universality.

The second term is equal to $f({{M\sqd}\over{\L\sqd}})$, where
\ba\begin{array}{c}
f(x) \equiv \int \ll 0 \uu\infty du~
B\pr (u) \ln \lrd x + u \rrd \\\\
\end{array}\ea
We are interested in the coefficient of $\ln x$ in the
expansion of $f(x)$ as $x\to 0$.  To extract this coefficient,
we can differentiate under the integral sign:
\ba\begin{array}{c}
{{d ~f(x)}\over{d ~\ln x}}  = x f\pr (x)
\\\\
= \int \ll 0\uu\infty {{B\pr(u)~x~du}\over{u + x}}
\\\\
= \left . {{x~B(u)}\over{x + u}} \right | \ll 0\uu\infty
+ \int\ll 0\uu\infty ~{{B(u) ~x~du}\over{(u+x)\sqd}}
\\\\
= - 1 + \int \ll 0 \uu\infty {{B(x v)~dv}\over
{(v+1)\sqd}}
\end{array}\ea
In the limit $x \to 0$ the function $B(xv)$ approaches $1$
for any finite $v$, so we have
\ee{
- 1 + \int\ll 0\uu \infty {{dv}\over{(v+1)\sqd}} = 0
}
In fact, it is easy to check that the limit $\L\to \infty$ is
finite, and given by the integral
\ba\begin{array}{c}
I\ll 3 \equiv
\int \ll 0 \uu\infty ~B\pr (u) ~du~\ln u,
\end{array}\ea
giving a finite renormalization of the cosmological constant
which can be removed with a counterterm.

So we are left with
\ba\begin{array}{c}
\int \ll 0 \uu\infty~{{B(u) du}\over{x + u}},
\end{array}\ea
with $x = {{M\sqd}\over{\L\sqd}},$ as before.  Again, taking
$x{d\over{dx}}$ of this quantity and taking $x\to 0$,
we get
\ba\begin{array}{c}
- \int\ll 0 \uu\infty {1\over{(v + 1)\sqd}} = -1
\end{array}\ea
So the total $T\ll{-1}$ contribution of the $X,\l,\psx$
degrees of freedom to the effective action
\ba\begin{array}{c}
-{\hbar\over 2} \lsq
I\ll 3 + I\ll 4 - \ln \lrd {{M\sqd}\over{\L\sqd}} \rrd \rsq,
\end{array}\ea
where
\ee{
I\ll 4 \equiv \int\ll 0\uu\infty {{B(u) ~du}\over{u+1}}
}
in the limit $\L\to\infty$.

There are still nonzero contributions at the $T\ll 0$ and
$T\ll 1$ terms, but these are finite and so are easier
to evaluate.

\subsection{Bosonic and fermionic
$T\ll 0$ and $T\ll {1}$ contributions}

There are also contributions to the effective dilaton
coupling from
subleading terms in the Euler-Maclaurin expansion.  We
now evaluate the $T\ll 0$ and $T\ll{1}$ contributions.

\subsubsection{Bosonic $T\ll 0$ contribution }

Since $B(u)$ vanishes with all its derivatives at $u\to\infty$,
so $T\ll 0$ contribution to the logarithm of the bosonic
determinant is
\ba\begin{array}{c}
\hh \left . \lsq (2n+1) \ln\lrd {{M\sqd}\over{({\mu\pr})\sqd}}
+ {{n\sqd + n}\over{\rws\sqd ({\mu\pr})\sqd}} \rrd \cdot
B\lrd {{n\sqd + n}\over{\rws\sqd \L\sqd}}\rrd \rsq
 \right | \ll{n = 0} = \hh \ln \lrd {{M\sqd}\over{({\mu\pr})\sqd}}
\rrd
\end{array}\ea

$T\ll 0$ and higher terms are automatically
ultraviolet finite, so $\L$ cannot enter them.  Also, they will
depend on the regulator only through its derivatives
at the origin.  $T\ll 0$ contributions, in particular,
are independent of the regulator function altogether.

\subsubsection{Fermionic $T\ll 0$ contribution}

The
\ba\begin{array}{c}
\hh \left .
\lsq ~(2n + 2) ~\ln \lrd
{{M\sqd}\over{({\mu\pr}) \sqd}}
 + {{(n+1)\sqd} \over{\rws\sqd({\mu\pr})\sqd}}
 \rrd ~B\lrd {{(n+1)\sqd - \hh}\over{\rws\sqd\L\sqd}} \rrd \rsq
\right | \ll {n=0}\\\\
= 1 \cdot \ln \lrd {{M\sqd}\over{({\mu\pr})\sqd}}\rrd + o\lrd
{1\over{\rws\sqd}} \rrd
\end{array}\ea

\subsubsection{Bosonic and fermionic $T\ll 1$ contributions}

These contributions come by differentiating the expression for
the $n\uu{\rm{\underline{th}}}$ term by $n$ and evaluating
at zero.  Up to terms of order $\rws\uu{-2}$, both bosonic
and fermionic summands have the same $n$-derivative at
zero, namely
\ee{
2 \ln\lrd {{M\sqd}\over{({\mu\pr})\sqd}}\rrd.
}
Since bosonic and fermionic determinants contribute oppositely
to the effective action, these two contributions cancel.

\subsubsection{Summing up the effective dilaton contributions}

The contributions to the effective action are as follows:
\ba\begin{array}{c}
{\hbar\over 2} \lsq \ln \lrd {{M\sqd}\over{\L\sqd}} \rrd
- I\ll 3 - I\ll 4 + \hh \ln \lrd {{M\sqd}\over{({\mu\pr})\sqd}} \rrd
-  \ln\lrd {{M\sqd}\over{({\mu\pr})\sqd}} \rrd
\rsq = \lrd {\rm terms~which~vanish~as~}\L\to \infty \rrd
\nn
\end{array}\ea
So as the regulator is removed and setting $\hbar = 1$,
we are left with
\ba\begin{array}{c}
= {1\over 4} \ln \lrd {{M\sqd}\over{\m\sqd}} \rrd +
\lrd M-{\rm independent}\rrd .
\end{array}\ea

So, the effective action on the sphere is enhanced by a factor
of $\hh b X\uu 0 + \lrd X\uu 0-{\rm independent} \rrd$,
which we interpret as being equal to $+2 \Delta\ll
{\rm one-point} \Phi$, since
we are working on the sphere.  The notation is intended to
convey the idea that this renormalization comes
from the renormalized one-point function for $X\uu 0$
(extracting the subleading term in volume on a finite-sized
sphere).

\appendix{Wavefunction renormalization of $X\uu 0$}

There is another contribution to the dilaton coupling renormalization
which comes from the renormalization of the
\it two-point \rm function for $X\uu 0$, which we shall call
$\Delta\ll{\rm two-point} \Phi$.

The origin of this term is as follows.  One effect of integrating
out the $X,\l,\psx$ degrees of freedom is that $X\uu 0$
receives a wavefunction renormalization, which as we shall see
is proportional to $b\sqd$.  Let us say
the (Euclidean) kinetic term receives an effective
contribution ${1\over{4\pi}} (c\ll 1  b\sqd)
(\vp \xx)\sqd$, where $c\ll 1$ is some numerical constant which
will be calculated from a loop diagram.  Then in order to
re-canonicalize the field $\xx$ at large $\xx$ we must
rescale $\xx$ by a factor $(1 + c\ll 1 b\sqd)\uu{-\hh}
\sim 1 - \hh c\ll 1 b\sqd + o (b\uu 4)$.
As a result, the $\qe X\uu 0 {\cal R}$ term in the original
bare lagrangian gets scaled by that factor and the result
is a term equal to
\bbb
-\hh c\ll 1 b\sqd \qe \cdot{1\over{4\pi}}\int {d\sqd\s}~\sqrt{g}~
{\cal R}\up 2{\tilde{\xx}},
\eee
where $\tilde{\xx}$ is the field $\xx$, rescaled so that its
loop-corrected effective kinetic term has
canonical normalization.

In the
small-b branch of the large-n limit,
the original dilaton gradient $\qe$ is of order $n\uu{+\hh}$
and $b$ is of order $n\uu{-\hh}$.  So the
'two-point' renormalization $\delta \qe = - \hh c\ll 2 b\sqd \qe$
is of order $n\uu{-\hh}$, just like the 'one-point'
contribution $\delta \qe = {1\over 4} b$.  We must take
this effect into account as well.

In the following section, we will not bother to make the
volume finite, since the term we are interested in scales as
$\rws\sqd$; it is an extensive term.

\subsection{Effective kinetic term from effective action}

The effective action coming from the loop of massive particles
we can calculate in terms of logarithms of determinants.  In
this subsection we will write the expression for the effective
kinetic term in terms of the effective action.

If our effective action is of the form
\ee{
{\cal S} = \int~d\sqd\s ~\k~(\vp \xx)\sqd,
}
then we can extract $\k$ by setting
\ee{
\xx\equiv w + \e~\exp{ik\s} + \e^* \exp{-ik\s},
}
where $w$ is the constant background value of $\xx$,
and taking derivatives of the action with respect to $k$ and
$\e\up{1,2}$.  Plug in this value for $\xx$ and
the action is
\ba\begin{array}{c}
\k\lsq (\e \cdot i\vec {k} )(\e^* \cdot (-i\vec{k})) \cdot 2
\rsq \lsq V\ll{WS} \rsq
\\\\
= 2 \k~V\ll{WS} \e \e^* {\vec k} \sqd ,
\end{array}\ea
where $V\ll{WS}$ is the volume of the worldsheet
$\int d\sqd \s$.

Take ${\d\over{\e} } {\d\over{\e^*}}$ of this
and get $2\k V\ll{WS} \vec{k}\sqd$.

Take $\hh {\d\over{\d k\sqd}}
= {1\over 8} {\d\over{\d k\uu a}} {\d\over{\d k\ll a}}$ of
this and get $\k~\vws$.  So
\ba\begin{array}{c}
\k = \left .
{1\over {8~\vws}} {\d\over{\d k\uu a}} {\d\over{\d k\ll a}}
{\d\over{\d\e} } {\d\over{\d\e^*}} {\cal S}\lsq w +
 \e~\exp{ik\s} + \e^* \exp{-ik\s} \rsq
\right |\ll{\e = k = 0}
\\\\
= \lno
{1\over{2~\vws}}
 {\d\over{\d k\sqd}} {\d\over{\d\e} } {\d\over{\d\e^*}}
{\cal S}\lsq w +
 \e~\exp{ik\s} + \e^* \exp{-ik\s} \rsq
\right |\ll{\e = k = 0} \nn
\end{array}\ea
So now let us evaluate the bosonic and fermionic
determinants, evaluated with the background value
\ee{
\xx = w +
 \e~ \exp{ik\s} + \e^* \exp{-ik\s}
}
Since we are aiming to extract only the 'extensive' contribution,
we will take only the $T\ll{-1}$ term, and the volume of
the worldsheet will enter only through the formula
\ee{
{\rm Tr}~ f(-i\pp\ll a) = {{V\ll{WS}}\over{(2\pi)\sqd}}
\int d\sqd p~ f(p\ll a)
}

\subsection{Bosonic contribution}

Defining $M(\s)\equiv \m~\exp{b\xx(\s)}$ and
$M\lo 0 \equiv \m~\exp{b\xx\lo 0}$,
the log of the bosonic determinant is
\ba\begin{array}{c}
{\rm Tr} ~\ln~\lsq {1\over{({\mu\pr})\sqd}}
\lrd (-i\pp)\sqd + M(\s)\sqd\rrd \rsq
 \\\\
= {\rm Tr} ~\ln~\lsq {1\over{({\mu\pr})\sqd}}
\lrd (-i\pp)\sqd + M\sqd \rrd \rsq +
 {\rm Tr} ~ \lsq \lrd (-i\pp)\sqd + M\sqd \rrd \uu{-1}
~\d M\sqd (\s) \rsq
\\\\
 - \hh {\rm Tr} ~ \lsq \lrd (-i\pp)\sqd + M\sqd \rrd \uu{-1}
~\d M\sqd (\s) ~
\lrd (-i\pp)\sqd + M\sqd \rrd \uu{-1}
~\d M\sqd (\s)
\rsq + o\lrd \d M\sqd(\s)\rrd \uu 3
\end{array}\ea

Only the term of order $o \lrd \d M\sqd (\s) \rrd\uu 2$
will give a nonvanishing $\e \e^*$ piece.  Using
$\d M\sqd = 2 b M\sqd \d\xx$, we find this
piece is equal to
\ba\begin{array}{c}
- 4 b\sqd M\uu 4 ~\e \e^*~
{\rm Tr} ~ \lsq \lrd (-i\pp)\sqd + M\sqd \rrd \uu{-1}
~\exp{i k\s}  ~
\lrd (-i\pp)\sqd + M\sqd \rrd \uu{-1}
~\exp{- i k\s}
\rsq \\\\
= - 4b\sqd M\uu 4 \e \e^* ~{\rm Tr} ~ \lsq \lrd
(-i\pp + k )\sqd + M\sqd \rrd \uu{-1}
\lrd (-i\pp)\sqd + M\sqd \rrd \uu{-1} \rsq
\\\\
= - {{ 4b\sqd M\uu 4
\e \e^* V\ll{WS}}\over{4\pi\sqd}}
 \int d\sqd p ~ \lsq \lrd
(p + k )\sqd + M\sqd \rrd \uu{-1}
\lrd p\sqd + M\sqd \rrd \uu{-1} \rsq
\end{array}\ea

We are only interested in the piece which is quadratic in $k$,
which is
\ba\begin{array}{c}
- {{ 4b\sqd M\uu 4 \e \e^* V\ll{WS}}\over{4\pi\sqd}}
\int d\sqd p ~ \lsq
{{-k\sqd}\over{(p\sqd + M\sqd)\uu 3}} + {{4 k\ll a k\ll b
p\uu a p\uu b}\over{(p\sqd + M\sqd)\uu 4}} \rsq \\\\
= + {1\over{6\pi}} \cdot  b\sqd \cdot V\ll{WS} \cdot \e
\e^* k\sqd
\end{array}\ea

To get the corresponding contribution to the effective
action, multiply by $+{\hbar\over 2}$, to get
\ee{
\Delta {\cal S}\ll{{\rm bos.}}
= + {{1}\over{12\pi}} \cdot \hbar\cdot
 b\sqd \cdot V\ll{WS} \cdot \e
\e^* k\sqd.
}
\subsection{Fermionic contribution}

The fermionic calculation is similar.  The determinant is
\ba\begin{array}{c}
{\rm Tr}~\ln~\lsq \cm(M + \d M) \rsq,
\end{array}\ea
where $\cm \equiv \lsq \G\uu i  (- i\pp\ll i) + i M(\s) \rsq
= \lsq (\phh\cdot \G) + i M(\s) \rsq$.
This operator is not Hermitean, but its determinant is real.

For constant $M$, its inverse
\ee{
\dh(\phh,M) =
\cm\uu{-1}(\phh,M) \equiv (\phh\sqd + M\sqd)\uu{-1}
((\phh\cdot \G) - i M)
}
defines the propagator.

The piece of the log of the determinant with two fluctuations is
\ba\begin{array}{c}
= - \hh ~ {\rm Tr}~\lsq \dh (\phh,M)
 \cdot  (-i\d M) \cdot
 \dh(\phh,M)
\cdot (-i\d M) \rsq
\\\\
= + b\sqd M\lo 0 \sqd |\e|\sqd~
 {\rm Tr}~\lsq \dh (\phh,M)
 \cdot  \exp{i k\s} \cdot
 \dh(\phh,M)
\cdot \exp{- i k\s}  \rsq
\\\\
= b\sqd M\lo 0 \sqd |\e|\sqd
 ~ {\rm Tr}~\lsq \dh (\phh + k ,M)
 \cdot
 \dh(\phh ,M)
  \rsq
\\\\
= b\sqd M\lo 0 \sqd |\e|\sqd
 {{V\ll{WS}}\over{4\pi\sqd}}\cdot \int d\sqd p
~{\rm tr}
\lsq \dh (p + k ,M)
 \cdot
 \dh(p ,M)
  \rsq
\\\\
=  + 2(b~M~\e)(b~M~\e^*)
\cdot \hh  \cdot {{V\ll{WS}}\over{4\pi\sqd}}\cdot I(k\sqd), \nn
\end{array}\ea
where
\ba\begin{array}{c}
I(k\sqd) \equiv
 \int d\sqd p ~{\rm tr}
\lsq \lrd \G \cdot (p+k) + i M\rrd \uu{-1} \cdot
\lrd \G \cdot p + i M \rrd \uu{-1}
  \rsq
\end{array}\ea
We wish to extract the order $k\sqd$ piece $I\pr(0) =
\left . I\pr (k\sqd) \right |\ll{k = 0} =
\left .
{1\over 4} {\d\over{\d k\uu a}}
{\d\over{\d k \ll a}} ~I(k\sqd)\right |\ll{k = 0}
$ of the amplitude $I(k\sqd)$.  Differentiating under
the integral sign, we get
\ba\begin{array}{c}
{\d\over{\d k\uu a}} I(k\sqd) = -  \int d\sqd p ~{\rm tr}
\lsq \dh(p+k,M)
\G\uu a  \dh(p+k,M)
 \dh(p,M)
  \rsq
\nn\\
\end{array}\ea
and
\ba\begin{array}{c}
{\d\over{\d k\uu a}}
{\d\over{\d k\ll a}}
 I(k\sqd) = + 2 \int d\sqd p ~{\rm tr}
 \lsq \dh(p+k,M) \G\ll a   \dh(p+k,M) \G\uu a \dh(p+k,M)
\dh(p,M)
 \rsq \nn
\end{array}\ea
so
\ba\begin{array}{c}
I\pr (0) = \lno {1\over 4} {\d\over{\d k\uu a}}
{\d\over{\d k\ll a}}
 I(k\sqd) \rba \ll {k = 0}
\nn\\\\
= \hh \int d\sqd p ~{\rm tr} ~
\lsq \dh  \uu 3 (p,M) \G\uu a \dh(p,M) \G\ll a \rsq
\end{array}\ea

Using the fact that
\ba\begin{array}{c}
\tr \lsq \dh\uu 3 (p,M) \G\uu a \dh(p,M) \G\ll a \rsq
= 4 M\sqd {{M\sqd - 3 p\sqd}\over{(p\sqd + M\sqd)\uu 4}}
\end{array}\ea
and we conclude that
\ba\begin{array}{c}
I\pr (0) = \hhc 4 M\sqd \int ~d\sqd p~{{M \sqd  - 3 p\sqd }\over{
(p\sqd + M\sqd)\uu 4 }}
\nn
\end{array}\ea
and
\ba\begin{array}{c}
\lno \ln~{\rm Det}(A)\rba\ll{o(\e\st\e~k\sqd)}
= - {1\over{12\pi}}  b\sqd \vws \e \e^* k\sqd.
\end{array}\ea

To get the corresponding contribution to the action,
multiply by $-{\hbar\over 2}$, to get
\ba\begin{array}{c}
\Delta {\cal S}\ll{\rm ferm.} = + {\hbar\over{24\pi}}
b\sqd \vws \e \e^* k\sqd.
\end{array}\ea


\newpage
\begin{thebibliography}{99}


\small
\parskip=0pt plus 2pt
\bibitem{Hellerman}
S.~Hellerman,
``On the Landscape of Superstring Theory in D $>$ 10,''
arXiv:hep-th/0405041.

\bibitem{Strominger}
A.~Strominger and T.~Takayanagi,
``Correlators in timelike bulk Liouville theory,''
Adv.\ Theor.\ Math.\ Phys.\  {\bf 7}, 369 (2003)
[arXiv:hep-th/0303221].

\bibitem{deAlwis}
S.~P.~de Alwis, J.~Polchinski and R.~Schimmrigk,
``Heterotic Strings With Tree Level Cosmological Constant,''
Phys.\ Lett.\ B {\bf 218}, 449 (1989).

\bibitem{Chamseddine}
A.~H.~Chamseddine,
``A Study of noncritical strings in arbitrary dimensions,''
Nucl.\ Phys.\ B {\bf 368}, 98 (1992).


\bibitem{joebook2}
J.~Polchinski,
``String Theory. Vol. 2: Superstring Theory And Beyond,''~
(Cambridge University Press, 1998); correction to eq. (12.1.34b)
at http://theory.itp.ucsb.edu/~joep/errata.html\#c12.


\bibitem{sw}
S.~Sugimoto,
``Anomaly cancellations in type I D9-D9-bar system and the USp(32)  string
Prog.\ Theor.\ Phys.\  {\bf 102}, 685 (1999)
[arXiv:hep-th/9905159] ;~
J.~H.~Schwarz and E.~Witten,
JHEP {\bf 0103}, 032 (2001)
[arXiv:hep-th/0103099].
%

\bibitem{Myers}
R.~C.~Myers,
``New Dimensions For Old Strings,''
Phys.\ Lett.\ B {\bf 199}, 371 (1987).


\bibitem{lt1}
T.~L.~Curtright and C.~B.~Thorn,
Phys.\ Rev.\ Lett.\  {\bf 48}, 1309 (1982)
[Erratum-ibid.\  {\bf 48}, 1768 (1982)].

\bibitem{lt2}
T.~Fukuda and K.~Hosomichi,
``Three-point functions in sine-Liouville theory,''
JHEP {\bf 0109}, 003 (2001)
[arXiv:hep-th/0105217].

\bibitem{lt3}
N.~Seiberg,
``Notes On Quantum Liouville Theory And Quantum Gravity,''
Prog.\ Theor.\ Phys.\ Suppl.\  {\bf 102}, 319 (1990).

\bibitem{lt4}
A.~M.~Polyakov,
Phys.\ Lett.\ B {\bf 103}, 211 (1981).

\bibitem{lt5}
E.~D'Hoker,
Phys.\ Rev.\ D {\bf 28}, 1346 (1983).

\bibitem{lt6}
D.~Kutasov and N.~Seiberg,
Phys.\ Lett.\ B {\bf 251}, 67 (1990).

\bibitem{lt7}
H.~Dorn and H.~J.~Otto,
Nucl.\ Phys.\ B {\bf 429}, 375 (1994)
[arXiv:hep-th/9403141].

\bibitem{lt8}
M.~Goulian and M.~Li,
``Correlation Functions In Liouville Theory,''
Phys.\ Rev.\ Lett.\  {\bf 66}, 2051 (1991).

\bibitem{lt9}
J.~Teschner,
Phys.\ Lett.\ B {\bf 363}, 65 (1995)
[arXiv:hep-th/9507109].

\bibitem{eigenvalues}
R.~Camporesi and A.~Higuchi, ``On the Eigen functions of the Dirac
operator on spheres and real hyperbolic spaces,''
arXiv:gr-qc/9505009.



\end{thebibliography}
\end{document}